\documentclass[
reprint,
superscriptaddress,
amsmath,
amssymb,
aps,
longbibliography,
pra,
showpacs,
floatfix
]{revtex4-2}

\usepackage{hyperref}
\usepackage{graphicx}
\usepackage{xcolor} 

\usepackage{tabularx}
\usepackage{multirow}
\usepackage{dcolumn} 
\newcolumntype{d}[1]{D{.}{.}{#1}}
\newcolumntype{L}[1]{>{\raggedright\arraybackslash}p{#1}}
\newcolumntype{C}[1]{>{\centering\arraybackslash}p{#1}}
\newcolumntype{R}[1]{>{\raggedleft\arraybackslash}p{#1}}

\usepackage{braket}

\usepackage{comment}
\usepackage{enumitem}
\usepackage{soul}

\hypersetup{colorlinks=true, 
    linkcolor={blue},
    citecolor={blue}, 
    urlcolor={blue}
}

\begin{document}

\title{Phonon polariton Hall effect}

\author{Omer Yaniv}
\email{yaniv@mail.tau.ac.il}
\affiliation{School of Physics and Astronomy, Tel Aviv University, Tel Aviv 6997801, Israel}

\author{Dominik~M.\ Juraschek}
\email{d.m.juraschek@tue.nl}
\affiliation{School of Physics and Astronomy, Tel Aviv University, Tel Aviv 6997801, Israel}
\affiliation{Department of Applied Physics and Science Education, Eindhoven University of Technology, 5612 AP Eindhoven, Netherlands}

\date{\today}


\begin{abstract}
The phonon Hall effect conventionally describes the generation of a transverse heat current in an applied magnetic field. In this work, we extend the effect to hybrid light–matter excitations and demonstrate theoretically that phonon polaritons, formed by coupling optical phonons with terahertz radiation,  support transverse energy flow when coherently driven in an applied magnetic field. Using the example of PbTe, which exhibits strongly coupled phonon polaritons, we show that the magnetic field splits the phonon-polariton branches into left and right-handed circular polarization, obtaining unequal group velocities. We derive the energy current operators for propagating phonon polaritons and show how their mixed phononic-photonic nature enables controllable transverse phonon-polariton transport in the terahertz regime. Our results demonstrate bending of light through a phonon polariton Hall effect, which provides a route towards terahertz polaritonic devices.
\end{abstract}

\maketitle


\section{Introduction}
The phonon Hall effect describes the generation of a transverse heat current in response to a longitudinal temperature gradient in an externally applied magnetic field \cite{strohm:2005,sheng:2006}. It has been observed in semiconducting and insulating materials, and is understood in terms of the non-vanishing Berry curvature of phonon bands induced by the coupling between a magnetic field and the angular momentum of phonons \cite{sheng:2006,Inyushkin2007,Kagan2008,Wang2009,zhang2010topological,Zhang2011,agarwalla2011phonon,Qin2012,mori2014origin}. In recent years, the phenomenon has attracted increasing attention, as giant Hall angles have been reported across a broad range of materials, and the Hall conductivity itself has been found to follow a universal scaling behavior \cite{ideue2017giant,Grissonnanche2019,Grissonnanche2020,Li2020,yang2020universal,Guo2021phononHall,chen2022large,Mangeolle2022,Sun2022,Li2023phononHall,Flebus2023,Chen2024,Jin2024_PHE,Oh2024}. 

Because the phonon Hall effect is a type of thermal Hall effect, all studies thus-far focused on heat currents carried by thermally populated acoustic phonons initiated by a temperature gradient. The microscopic origins of symmetry breaking and phonon band geometry are not limited to the acoustic branches however and should equally apply to optical phonons. Owing to their strong coupling to electromagnetic fields, well-defined polarization, and suitability for selective excitation, optical phonons provide a promising platform for transverse phonon transport under coherent driving conditions.

In this work, we extend the concept of the phonon Hall effect to propagating phonon polaritons: hybrid quasiparticles formed by the interaction of optical phonons with terahertz radiation. Phonon polaritons inherit at the same time a finite group velocity from the photon part and the band geometry of the eigenvectors from the phonon part. The phonon-polariton branches are formed by the transverse optical phonons, which split in an applied magnetic field according to the phonon Zeeman effect \cite{schaack:1976,thalmeier1978magnetic,juraschek2:2017,Juraschek2019,Cheng2020,Baydin2022,Hernandez2023}. We show that this splitting in combination with the longitudinal propagation induces a transverse phonon-polariton current, which we call phonon polariton Hall effect, enabling a pathway to controlling light propagation at terahertz frequencies.


\begin{figure}[b]
\centering
\includegraphics[width=1\linewidth]{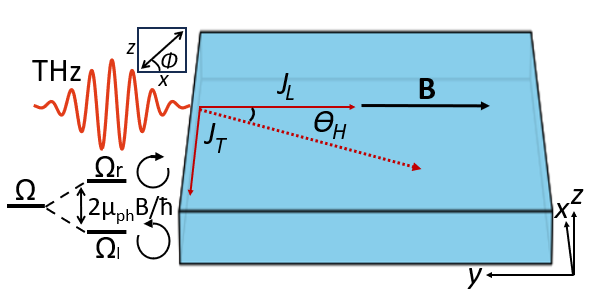}
\caption{Phonon polariton Hall effect. A thin slab supports phonon propagation in the $xy$ plane while preserving degeneracy of the transverse optical phonons. A terahertz pulse linearly polarized at $\phi=45^\circ$ in the $xz$ plane and propagating in the $y$ direction launches a longitudinal phonon-polariton current, $J_L$. A magnetic field, $\mathbf{B}$, applied along the $y$ axis induces Zeeman splitting between between left- and right-handed circularly polarized phonons, $\Omega_l$ and $\Omega_r$, inducing a transverse phonon-polariton current, $J_T$.
}
\label{fig:setup}
\end{figure}


\section{Phonon-polariton branches and Zeeman splitting}

We investigate the example of lead telluride (PbTe), a narrow-bandgap semiconductor that has been shown to exhibit strongly coupled phonon polaritons~\cite{baydin2025terahertz}, as well as a strong phonon Zeeman splitting of axial phonons up to 2\% when doped with tin (Sn) \cite{Baydin2022,Hernandez2023}. We consider a thin crystalline sample in the geometry illustrated in Fig.~\ref{fig:setup}. A terahertz pulse incident along the $y$ axis and linearly polarized at $\phi=45^\circ$ in the $xz$ plane irradiates the material and generates a phonon polariton propagating along the $y$ direction. The sample is chosen thin enough to prevent out-of-plane ($z$-axis) propagation, but thick enough to preserve the degeneracy of the transverse optical (TO) phonon branches in the $xz$ plane. At the same time, a magnetic field is applied parallel to the propagation direction, $\mathbf{B}=B\hat{y}$, which splits the degenerate TO branches into left-handed and right-handed circular polarization, which can be considered chiral phonon polaritons \cite{thalmeier1978magnetic,Biggs2025}. The splitting arises from the coupling of the magnetic field to the phonon magnetic moment, which has previously been found in nonmagnetic materials \cite{juraschek2:2017,Juraschek2019,Cheng2020,Geilhufe2021,Xiao2021,Ren2021,Baydin2022,Hernandez2023,Geilhufe2023,Zhang2023_BLG,Shabala2024,Klebl2024,Mustafa2025,Chen2025_gaugetheory,Paiva2025}. This magnetic moment originates from the circular motion of ions in the lattice and and can be enhanced by electron-phonon interactions, ranging between a nuclear magneton ($\mu_N$) and a Bohr magneton ($\mu_B$). We will in the following derive the chiral phonon-polariton dispersion.

The classical phonon-polariton dispersion relation in cubic systems is given by \cite{Horton1974}
\begin{equation}\label{eq:classical_dispersion}
\omega_{\mathbf{k},\pm}^2 =  \frac{1}{2} \Big( \Omega_L^2 + \omega^2_{\nu}(\mathbf{k}) 
\pm \sqrt{(\Omega_L^2 + \omega^2_{\nu}(\mathbf{k}))^2
- 4 \Omega_T^2 \omega^2_{\nu}(\mathbf{k})} \Big),
\end{equation}
where $\omega_L = \sqrt{\omega_T^2 + \frac{Z^2}{\varepsilon_0 \varepsilon_\infty V_c}}$ is the longitudinal optical (LO) phonon frequency and $\omega_{\nu}(\mathbf{k}) = \frac{c k}{\sqrt{\varepsilon_\infty}}$ is the photon dispersion in the material. $Z$ is the mode effective charge, $V_c$ the unit cell volume, $
\varepsilon_\infty$ the high-frequency dielectric constant, and $\varepsilon_0$ the vacuum permittivity. The two phonon-polariton branches $\omega_{\mathbf{k},\pm}$ interpolate between the bare photon and phonon dispersions. Because of this hybridization, any perturbation acting on the phonon part is directly inherited by the polaritons. For small wavevectors, the frequencies of the TO phonons can be considered constant, $\Omega_T(\mathbf{k})\equiv\Omega_T$.

In second quantization, the harmonic Hamiltonian reads
\begin{equation}
H_0=\sum_{\mathbf{k},\sigma=l,r} \hbar \Omega_T
\Big(a^{\dagger}_{\mathbf{k},\sigma} a_{\mathbf{k},\sigma}+\tfrac12\Big).
\end{equation}
The coupling of the magnetic field to the phonon magnetic moment can be written as
\begin{align}
H_Z
=-\sqrt{\frac{\Omega_{\mathbf{k},\sigma'}}{\Omega_{\mathbf{k},\sigma}}}\mu_{\mathrm{ph}}\sum_{\mathbf{k},\sigma}
\mathbf{B}\cdot(i\mathbf \epsilon_{\mathbf{k},\sigma}^*\times \epsilon_{\mathbf{k},\sigma'})\;
a^{\dagger}_{\mathbf{k},\sigma} a_{\mathbf{k},\sigma'},
\end{align}
where $\epsilon_{\mathbf{k},\sigma}$ are the eigenvectors of the doubly degenerate TO phonons in circular basis, dependent on the wavevector $\mathbf{k}$ and circular polarization $\sigma \in \{l,r\}$. They satisfy $i \epsilon_{\mathbf{k},\sigma}^*\times \epsilon_{\mathbf{k},\sigma} = s_\sigma$, where $s_l=+1,\; s_r=-1$. Treating $H_Z$ to first order shifts the TO frequencies by $\Delta \Omega_{\sigma}=\langle 1_{\mathbf k,\sigma}|H_Z|1_{\mathbf k,\sigma}\rangle/\hbar$, yielding $\Omega_\sigma=\Omega_T-s_\sigma\frac{\mu_{\mathrm{ph}} B}{\hbar}$.

Substituting the modified transverse phonon frequencies $\Omega_\sigma$ into Eq.~\eqref{eq:classical_dispersion} splits the degenerate upper and lower phonon-polariton branches, respectively, yielding four nondegenerate branches in total \cite{thalmeier1978magnetic}. To quantify the light–matter hybridization and capture the evolution of the mixed character in the polariton states, we model the interaction using a polarization-resolved quantum avoided crossing Hamiltonian,
\begin{equation}
H
= \hbar\sum_{\mathbf k}
\begin{pmatrix}
a_{\mathbf{k},\sigma}^\dagger & c_{\mathbf{k}}^\dagger
\end{pmatrix}
\begin{pmatrix}
\Omega_{\sigma} & g(\mathbf{k}) \\
g^*(\mathbf{k}) & \omega_\nu(\mathbf{k})
\end{pmatrix}
\begin{pmatrix}
a_{\mathbf{k},\sigma} \\
c_{\mathbf{k}}
\end{pmatrix},
\end{equation}
where $a_{\mathbf k,\sigma}^\dagger$, $a_{\mathbf k,\sigma}$ are the creation and annihilation operators of the TO phonons and $c_{\mathbf k}^\dagger$, $c_{\mathbf k}$ those of the photons. Here, $g(\mathbf k)$ denotes the momentum-dependent photon–phonon coupling strength. Diagonalizing this Hamiltonian (see Supplemental Material for details \cite{SUPP}) yields the phonon-polariton eigenfrequencies,
\begin{equation}
\omega_{\mathbf{k},\sigma,\pm} = \frac{1}{2} \Big( \Omega_{\sigma} + \omega_{\nu}(\mathbf{k}) \pm \sqrt{\left(\Omega_{\sigma} - \omega_{\nu}(\mathbf{k})\right)^2 + 4|g|^2} \Big),
\label{eq:quantum_dispersion}
\end{equation}
and the phonon fraction of each phonon-polariton branch quantifies the relative weight of the vibrational component,
\begin{equation}
F_{\mathbf{k},\sigma,\pm} = \frac{1}{1 + \left(\frac{\omega_{\mathbf{k},\sigma,\pm} - \Omega_{\sigma}}{|g(\mathbf{k})|}\right)^2},
\label{eq:phonon_fraction}
\end{equation}
which provides a momentum-resolved measure of the phononic and photonic character of each phonon-polariton branch. By matching the classical dispersion in Eq.~\eqref{eq:classical_dispersion} to the quantum model in Eq.~\eqref{eq:quantum_dispersion} for each wavevector, we extract $g(\mathbf{k})$ and analyze the phonon content and transport properties of the upper and lower phonon-polariton branches in the presence of an external magnetic field.
\begin{figure*}[t]
    \centering
    \includegraphics[width=0.8\textwidth]{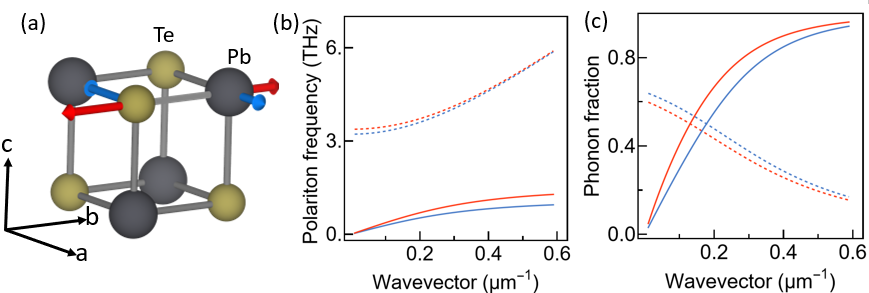}
    \caption{Phonon polaritons in PbTe. 
    (a) Cubic unit cell with TO phonon eigenvectors indicated as blue and red arrows.
    (b)~Dispersion relations for the upper (dashed) and lower (solid) phonon-polariton branches. Blue and red curves correspond to left-handed and right-handed circularly polarized phonons, respectively. The Zeeman splitting in an applied magnetic field lifts the degeneracy of the phonon-polariton branches, resulting in different group velocities for left- and right-handed circular polarizations.
    (c) Phonon fraction for the upper and lower phonon-polariton branches. For clarity of visualization, we use a splitting of $\Delta\Omega_\sigma=0.2$~THz, which is an order of magnitude larger than the experimentally measured values \cite{Baydin2022,Hernandez2023}.}
    \label{fig:polariton_pbte_cecl3}
\end{figure*}

We quantify the effect for the example of PbTe, a centrosymmetric semiconductor for which a strong phonon Zeeman effect has been measured \cite{Baydin2022,Hernandez2023}. We compute the phonon eigenfrequencies, eigenvectors, and Born effective charge tensors using density functional theory as implemented in VASP via the frozen-phonon method using the phonopy package ~\cite{kresse:1996,kresse2:1996,Togo2015}. We use the default projector augmented wave (PAW) pseudopotentials for each atom and converge the Hellmann-Feynman forces to 50~$\mu$eV/\AA. A plane-wave energy cutoff of 650~eV and a $2\times2\times2$ $\Gamma$-centered Monkhorst-Pack $k$-point mesh are used to sample the Brillouin zone~\cite{Monkhorst/Pack:1976}, using a $5\times5\times5$ supercell for the phononic properties. For the exchange-correlation functional, we employ the PBEsol variant of the generalized gradient approximation (GGA)~\cite{csonka:2009}. At finite wavevector, the triply degenerate infrared-active phonons in PbTe splits into a degenerate TO phonon polarized in the $xz$-plane with frequency $\Omega_T/2\pi = 1.31$~THz, and one LO phonon polarized along the $y$-axis with frequency $\Omega_L /2\pi = 3.25$~THz. The eigenvectors of the TO phonon within the cubic unit cell of PbTe are illustrated in Fig~\ref{fig:polariton_pbte_cecl3}(a). Our calculations yield a mode effective charge of $Z = 0.66~e/\sqrt{u}$, a high-frequency dielectric constant of $\varepsilon_\infty = 31.6$, and a unit-cell volume of $V_c = 66.87$~\AA$^3$.

Based on these results, we calculate the polariton dispersion and extract the phonon fraction of the hybrid branches. In Fig.~\ref{fig:polariton_pbte_cecl3}(b) and (c), we show the polariton dispersion and phonon fractions for PbTe, calculated for a Zeeman splitting of $\Delta \Omega_{\sigma}=0.2$~THz. This value is about an order of magnitude larger than the experimentally measured splittings \cite{Baydin2022,Hernandez2023} and is chosen here only to clearly visualize the lifting of the degeneracy between the left- and right-handed circularly polarized branches. Near a wavevector of $k=0.15~\mu$m$^{-1}$, the phonon-polariton branches exhibit maximum hybridization, $F_{\mathbf{k},\sigma}\approx0.5$, corresponding to an approximate 50/50 admixture of photonic and phononic components. In this regime, the phonon polariton inherits both the finite group velocity of the photon and the Zeeman coupling of the phonon. 


\section{Transverse phonon-polariton current}

We now show how time-reversal symmetry breaking by the applied magnetic field generates a transverse phonon–polaritonic current. The effect is analogous to the phonon thermal Hall response, but here it arises from coherently excited phonon polaritons rather than thermally populated acoustic modes. We derived the forward-propagating phonon energy current operator in second quantization (for the full derivation see Supplemental Material \cite{SUPP}), which reads
\begin{align}
J_i &= \frac{\hbar}{2V_c}\sum_{
\mathbf{k}= k_y>0}\Bigg[
\sum_{\sigma} \omega_{\mathbf{k},\sigma}\partial_{k_i}\omega_{\mathbf{k},\sigma}\;
a^\dagger_{\mathbf{k},\sigma} a_{\mathbf{k},\sigma}
+\frac{1}{2}\sum_{\sigma\neq\sigma'}\label{eq:Jgeneral}\\
&
\sqrt{\frac{\omega_{\mathbf{k},\sigma'}}{\omega_{\mathbf{k},\sigma}}}
\big(\omega_{\mathbf{k},\sigma}^2-\omega_{\mathbf{k},\sigma'}^2\big)
\big(\epsilon^{\ast}_{\mathbf{k},\sigma}\cdot\partial_{k_i}\epsilon_{\mathbf{k},\sigma'}\big)
a^\dagger_{\mathbf{k},\sigma} a_{\mathbf{k},\sigma'}+  h.c.
\Bigg], \nonumber
\end{align}
where wavevectors are restricted to $\mathbf{k}=(0,k_y,0)$ with $k_y>0$ according to the setup in Fig.~\ref{fig:setup}.
To model the driven response, we consider a terahertz pulse linearly polarized at $\phi=45^\circ$ in the $xz$ plane that coherently excites a wavepacket propagating along the positive $y$ direction (Fig.~\ref{fig:setup}), producing equal populations of coherent left- and right-handed circularly polarized phonon-polariton states with a relative phase of $\pi/2$. Each circular branch is described by a coherent polariton state $\ket{\alpha_{\mathbf{k},\sigma}}$, described by the phonon-polariton annihilation operator, $p_{\mathbf{k},\sigma}$, as $p_{\mathbf{k},\sigma}\ket{\alpha_{\mathbf{k},\sigma}}=\alpha_{\mathbf{k},\sigma}\ket{\alpha_{\mathbf{k},\sigma}}$, where $|\alpha_{\mathbf{k},\sigma}|^2$ is the mean occupation number per unit cell. We can write $p_{\mathbf{k},\sigma}$ in terms of bare phonon and photon operators via the phonon fraction $F_{\mathbf{k},\sigma}$ (Eq.~\eqref{eq:phonon_fraction}) as $p_{\mathbf{k},\sigma}=\sqrt{F_{\mathbf{k},\sigma}}a_{\mathbf{k},\sigma}+\sqrt{1-F_{\mathbf{k},\sigma}}c_{\mathbf{k},\sigma}$, which satisfy the relations $\langle a^{\dagger}_{\mathbf{k},\sigma}a_{\mathbf{k},\sigma}\rangle=F_{\mathbf{k},\sigma}$ and $\langle a^{\dagger}_{\mathbf{k},l}a_{\mathbf{k},r}\rangle =e^{2i\phi}\sqrt{F_{\mathbf{k},l}F_{\mathbf{k},r}}\equiv\sqrt{F_{\mathbf{k},l}F_{\mathbf{k},r}}$ for $\phi=45^\circ$. 

\begin{figure*}[t]
    \centering
    \includegraphics[width=0.9\textwidth]{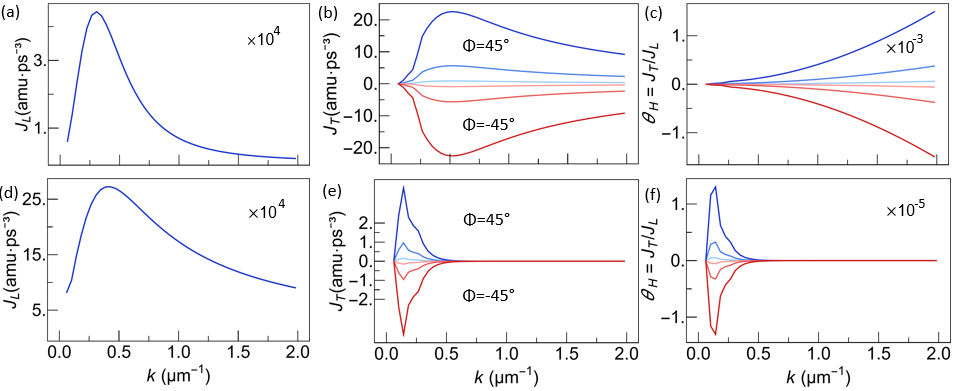}
  \caption{Calculated phonon-polariton currents in PbTe as a function of wavevector $k=k_y$. 
(a) Longitudinal phonon-polariton current, $J_L$, for the lower phonon-polariton branch. 
(b) Transverse current, $J_T$, for the lower branch. 
(c) Phonon polariton Hall angle $\theta_H = J_T / J_L$ for the lower branch.  
Panels (d) through (f) show the same quantities, respectively, for the upper phonon-polariton branch. For all plots of $J_T$ and $\theta_H$, we use three different values of the Zeeman splitting, $\Delta \Omega_{\sigma} \in \{0.004, 0.01, 0.02\}$~THz, indicated by light to dark shades. We further show two orientations of the pulse polarizationin the $xz$ plane, $\phi=\pm 45^\circ$, indicated by blue and red graphs, respectively.
}
\label{fig:transverse_current}
\end{figure*}

Inserting these relations into Eq.~\eqref{eq:Jgeneral} yields the expectation value of the phonon part of the phonon-polariton current operator,
\begin{align}
 \langle J_i \rangle = \frac{\hbar}{2V_c} \sum_{\mathbf{k}= k_y>0} \Bigg[ &
\sum_{\sigma}\omega_{\mathbf{k},\sigma}F_{\mathbf{k},\sigma} 
\partial_{k_i} \omega_{\mathbf{k},\sigma}  
\label{eq:energy_current} \\
& + \sum_{\sigma\neq\sigma'}\sqrt{\frac{\omega_{\mathbf{k},\sigma}}{\omega_{\mathbf{k},\sigma'}}}
\big(\omega_{\mathbf{k},\sigma'}^2-\omega_{\mathbf{k},\sigma}^2\big) \nonumber\\
& \times\Re \left\{ \big(\epsilon^{\ast}_{\mathbf{k},\sigma'}\cdot \partial_{k_i}\epsilon_{\mathbf{k},\sigma}\big)
\sqrt{F_{\mathbf{k},l}F_{\mathbf{k},\sigma}} i \right\} \Bigg] 
 \nonumber \\
& \equiv\frac{\hbar}{2V_c} \sum_{\mathbf{k}= k_y>0} \Bigg[J_L(\mathbf{k})+J_T(\mathbf{k}) \Bigg] \nonumber. 
\end{align}
Here, $J_L(\mathbf{k})$ and $J_T(\mathbf{k})$ are the wavevector-dependent longitudinal and transverse currents of the left- and right-handed circularly polarized phonon parts, produced by the first and second terms in Eq.~\eqref{eq:energy_current}, respectively. $J_T(\mathbf{k)}$ is proportional to the overlap $\epsilon^{\ast}_{\mathbf{k},l}\cdot \partial_{k_i}\epsilon_{\mathbf{k},r}$, where $\Re$ denotes the real part, which gives rise to the transverse current component. In order to reverse the direction of the transverse current, the sign of $\phi$ needs to be reversed, which in practice can be achieved by reorienting the polarization of the pulse by 90$^\circ$. To evaluate Eq.~\eqref{eq:energy_current}, we use the unperturbed eigenvectors of PbTe obtained from first principles. The magnetic field enters only through first-order corrections to the phonon frequencies, which are the source of the transverse current, while first-order corrections to the eigenvectors do not contribute, but are shown in Supplemental Material for completeness \cite{SUPP}.

We computed the phonon-polariton dispersion relations from Eq.~\eqref{eq:classical_dispersion} and \eqref{eq:quantum_dispersion}, and the phonon fraction $F_{\mathbf{k},\sigma}$ from Eq.~\eqref{eq:phonon_fraction}. The Zeeman splitting $\Delta \Omega_{\sigma}$ is chosen within the range of measured values reported in Refs.~\cite{Baydin2022,Hernandez2023}, ensuring consistency with experimentally accessible parameters. For PbTe, the splitting reaches about 2\% of the phonon eigenfrequency, while in PbSnTe it can be up to three times larger. We show the results in Fig.~\ref{fig:transverse_current} for magnetic fields pointing in both positive and negative $y$ directions, $\mathbf{B}=\pm\hat{y}$. Panels~(a) and~(d) show the longitudinal phonon-polariton current, $J_L$, for the lower and upper phonon-polariton branches, respectively. We find that significant propagation occurs only for wavevectors in the range of $k \le 2~\mu$m$^{-1}$, as $J_L$ rapidly vanishes for both branches for increasing wavevectors, because the group velocity of the lower branch and the phonon content of the upper branch each rapidly approach zero. 

Panels~(b) and~(e) display the transverse phonon-polariton current, $J_T$, for the upper and lower branches, respectively. $J_T$ reaches a maximum around $k \approx 0.5~\mu$m$^{-1}$ for the lower branch and $k \approx 0.15~\mu$m$^{-1}$ for the upper branch. For the upper branch, this corresponds to the region of maximum hybridization (Fig.~\ref{fig:polariton_pbte_cecl3}(c)), whereas for the lower branch, the maximum is located at larger wavevectors due to the increased Zeeman splitting as the branch becomes more phonon like. The corresponding phonon polariton Hall angles, $\theta_H = J_T/J_L$, are shown in panels~(c) and~(f). For the upper branch, the maximum of the Hall angle matches with the maximum of the transverse current. Intriguingly, the Hall angle $\theta_H$ for the lower phonon-polariton branch is ten times larger than that of the upper branch and grows monotonously with increasing wavevector. This behavior is not due to a divergence of the phonon polariton Hall current however, but is a consequence of the vanishing longitudinal current, corresponding to an evolution of the lower branch into surface phonon states that do no longer propagate into the bulk. The largest Hall response is therefore to be expected in the region between the maxima for $J_L$ and $J_T$, which lies around $k\approx 0.15~\mu$m$^{-1}$ for the upper branch and between $k\approx 0.4-0.5~\mu$m$^{-1}$ for the lower branch. In this regime, the phonon polariton Hall angles reach $\theta_H=10^{-4}$ for the lower branch and $10^{-5}$ for the upper branch, comparable with those found in the phonon thermall Hall effect~\cite{strohm:2005,Kagan2008,Mori2014,Li2023phononHall}.


\section{Discussion}

Our results demonstrate a phonon polariton Hall effect that enables manipulation of light propagation at terahertz frequencies. The effect is universal to all materials hosting infrared-active optical phonons carrying a phonon magnetic moment that can be driven with light and couple to an applied magnetic field. The phonon polariton Hall angle we compute is comparable to those found for acoustic phonons in the phonon thermal Hall effect and we expect it can further be enhanced in materials that also show giant phonon thermal Hall angles \cite{Grissonnanche2019,Grissonnanche2020} and phonon magnetic moments \cite{Chaudhary2024,Chen2025_gaugetheory}.

While the physical coupling mechanism between the applied magnetic field and the phonons is the same for the phonon polariton Hall and the phonon thermal Hall effects, there are some crucial differences. In the latter case, the splitting of the acoustic phonon branches leads to unequal thermal populations and therefore a transverse current that can be reversed by switching the direction of the magnetic field. In our case, the incident terahertz pulse generally populates both phonon-polariton branches equally and the transverse current arises from the unequal weights of the frequencies in Eq.~\eqref{eq:energy_current}. In order to reverse the transverse current, the polarization of the pulse therefore needs to be reoriented, which changes the relative phase between left- and right-handed circular polarizations. Varying the center frequency of the terahertz pulse in turn would provide a possibility to create unequal coherent population of the branches, leading to an additional contribution to the phonon polariton Hall current. Because coherent excitation produces nonequilibrium populations far exceeding thermal occupation, the net effect can be amplified well beyond what is possible with thermal gradients. 

Finally, we anticipate the effect to be measurable with state-of-the-art terahertz pump-probe setups that have in past years achieved to probe phonon-polariton dispersion relations and real-time propagation \cite{Henstridge2022,Lin2022,luo2024time,Sellati2025,Biggs2025} and offer a direct path to verifying our predictions.


\section*{Acknowledgments}

We thank Pooja Rani, Andrea Cavalleri, and Roberto Merlin for useful discussions. Calculations were performed on local HPC infrastructure at Tel Aviv University. This work was supported by the Israel Science Foundation (ISF) Grant No. 1077/23 and 1916/23.


%


\clearpage
\onecolumngrid

\setcounter{page}{1}
\setcounter{equation}{0}
\setcounter{figure}{0}
\setcounter{table}{0}
\setcounter{section}{0}
\makeatletter
\renewcommand{\theequation}{S\arabic{equation}}
\renewcommand{\thefigure}{S\arabic{figure}}
\renewcommand{\thetable}{S\arabic{table}}

\begin{center}
\textbf{\large Supplemental Material: Phonon polariton Hall effect}\\[0.4cm]
Omer Yaniv$^{1}$ and Dominik~M.~Juraschek$^{1,2}$\\[0.15cm]
\affiliation{}
$^1${\itshape{\small School of Physics and Astronomy, Tel Aviv University, Tel Aviv 6997801, Israel}}\\
$^2${\itshape{\small Department of Applied Physics and Science Education,\\ Eindhoven University of Technology, 5612 AP Eindhoven, Netherlands}}\\
\end{center}

\section*{Derivation of the phonon fraction}

\noindent{}The Hamiltonian of the hybrid phonon-polariton system can be written as 
\begin{equation}
    H=\hbar\sum_\mathbf{k}\Psi_{\mathbf k,\sigma}^\dagger M_{\mathbf k,\sigma}\Psi_{\mathbf k,\sigma},
\end{equation}
where
$\Psi_{\mathbf k,\sigma}
=
\begin{pmatrix}
a_{\mathbf k,\sigma}\\ c_{\mathbf k}
\end{pmatrix}$
and 
$
M_{\mathbf k,\sigma}
=
\begin{pmatrix}
\Omega_\sigma & g(\mathbf k)\\ g^*(\mathbf k) & \omega_\nu(\mathbf k)
\end{pmatrix}.
$
$\Omega_\sigma$ is the phonon frequency, $\omega_\nu(\mathbf k)$ is the photon frequency, $a_{\mathbf k,\sigma}^\dagger$, $a_{\mathbf k,\sigma}$ are the creation and annihilation operators of the TO phonons and $c_{\mathbf k}^\dagger$, $c_{\mathbf k}$ those of the photons. Here, $g(\mathbf k)$ denotes the momentum-dependent photon–phonon coupling strength. Since $M_{\mathbf k,\sigma}$ is Hermitian, it admits a unitary diagonalization
$U_{\mathbf k,\sigma}^\dagger M_{\mathbf k,\sigma}U_{\mathbf k,\sigma}
=\mathrm{diag}(\omega_{\mathbf k,\sigma,+},\omega_{\mathbf k,\sigma,-})$,
where the columns of $U_{\mathbf k,\sigma}$ are normalized eigenvectors
$\mathbf e_{\mathbf k,\sigma,\pm}=(u_{\mathbf k,\sigma,\pm},v_{\mathbf k,\sigma,\pm})^T$.
We define the phonon-polariton operators by the unitary rotation
\begin{equation}
\begin{pmatrix}
p_{\mathbf k,\sigma,+}\\[2pt] p_{\mathbf k,\sigma,-}
\end{pmatrix}
=
U_{\mathbf k,\sigma}^\dagger
\begin{pmatrix}
a_{\mathbf k,\sigma}\\[2pt] c_{\mathbf k}
\end{pmatrix}
=
\begin{pmatrix}
u_{\mathbf k,\sigma,+}^* & v_{\mathbf k,\sigma,+}^*\\
u_{\mathbf k,\sigma,-}^* & v_{\mathbf k,\sigma,-}^*
\end{pmatrix}
\begin{pmatrix}
a_{\mathbf k,\sigma}\\ c_{\mathbf k}
\end{pmatrix}.
\end{equation}
Because $U_{\mathbf k,\sigma}$ is unitary and $[a_{\mathbf k,\sigma},a_{\mathbf k,\sigma}^\dagger]=[c_{\mathbf k},c_{\mathbf k}^\dagger]=1$ with all other commutators being zero, the phonon-polariton operators obey bosonic commutation relations, $[p_{\mathbf k,\sigma,\mu},p_{\mathbf k',\sigma',\nu}^\dagger]=\delta_{\mathbf k\mathbf k'}\delta_{\sigma\sigma'}\delta_{\mu\nu}$.
In this basis, the Hamiltonian is diagonal and given by
\begin{equation}
 H=\hbar\sum_{\mathbf k,\sigma,\pm}
\omega_{\mathbf k,\sigma,\pm}
p_{\mathbf k,\sigma,\pm}^\dagger p_{\mathbf k,\sigma,\pm}.
\end{equation}
A one–polariton eigenstate can therefore be written as $|\psi_{\mathbf k,\sigma,\pm}\rangle=p_{\mathbf k,\sigma,\pm}^\dagger|0\rangle
= u_{\mathbf k,\sigma,\pm}^*a_{\mathbf k,\sigma}^\dagger|0\rangle
+ v_{\mathbf k,\sigma,\pm}^*c_{\mathbf k}^\dagger|0\rangle$,
where the probability to find a phonon quantum in this state is the squared modulus of the phonon component,
\(
F_{\mathbf k,\sigma,\pm}\equiv |u_{\mathbf k,\sigma,\pm}|^2.
\)
 The eigenvalue equation
\(
M_{\mathbf k,\sigma}\mathbf e_{\mathbf k,\sigma,\pm}
=\omega_{\mathbf k,\sigma,\pm}\mathbf e_{\mathbf k,\sigma,\pm}
\)
yields the ratio of the photon and phonon components,  
\(
\frac{v_{\mathbf k,\sigma,\pm}}{u_{\mathbf k,\sigma,\pm}}=
\frac{\omega_{\mathbf k,\sigma,\pm}-\Omega_\sigma}{|g(\mathbf k)|}.
\)
Normalizing $\mathbf e_{\mathbf k,\sigma,\pm}$ then yields
\begin{equation}
|u_{\mathbf k,\sigma,\pm}|^2
=\frac{1}{1+\big(\tfrac{\omega_{\mathbf k,\sigma,\pm}-\Omega_\sigma}{|g(\mathbf k)|}\big)^2}.
\end{equation}
Thus the phonon fraction becomes
\begin{equation}
F_{\mathbf{k},\sigma,\pm}=|u_{\mathbf{k},\sigma,\pm}|^2
=\frac{1}{1+\left(\frac{\omega_{\mathbf{k},\sigma,\pm}-\Omega_\sigma}{|g(\mathbf{k})|}\right)^2},
\end{equation}
which is Eq.~(6) in the main text.

\section*{Derivation of the phonon energy current}

\noindent{}We begin with the phonon energy current operator as in Ref.~[2], which is given by
\begin{equation}
    J_i = \frac{1}{2NV_c} \sum_{mn} (\mathbf{R}_m - \mathbf{R}_n)_i u_m^\alpha \Phi_{mn}^{\alpha\beta} \dot{u}_n^\beta .
    \label{eq:current}
\end{equation}
We next derive a quantum mechanical expression for the current operator. In order to arrive at a quantity defined in terms of phonons per unit cell, we normalize the displacement and velocity operators by $\sqrt{N}$. In second quantization, they become
\begin{align}
\frac{u_m^\alpha}{\sqrt{N}} &= \frac{1}{\sqrt{N}} \sum_{\mathbf{k},\sigma} \frac{1}{\sqrt{2M_{\alpha} \omega_{\mathbf{k},\sigma}}} \left(\frac{\tilde{a}_{\mathbf{k},\sigma}}{\sqrt{N}} \epsilon_{\mathbf{k},\sigma}^\alpha e^{i\mathbf{k} \cdot \mathbf{R}_m} + h.c. \right)=\frac{1}{\sqrt{N}} \sum_{\mathbf{k},\sigma} \frac{1}{\sqrt{2M_{\alpha} \omega_{\mathbf{k},\sigma}}} \left(a_{\mathbf{k},\sigma}\epsilon_{\mathbf{k},\sigma}^\alpha e^{i\mathbf{k} \cdot \mathbf{R}_m} + h.c. \right), \\
\frac{\dot{u}_n^\beta}{\sqrt{N}} &= \frac{-i}{\sqrt{N}} \sum_{\mathbf{k},\sigma} \sqrt{\frac{\omega_{\mathbf{k},\sigma}}{2M_{\beta}}} \left(-\frac{\tilde{a}_{\mathbf{k},\sigma}}{\sqrt{N}} \epsilon_{\mathbf{k},\sigma}^\beta e^{i\mathbf{k} \cdot \mathbf{R}_n} + h.c.\right)=\frac{-i}{\sqrt{N}} \sum_{\mathbf{k},\sigma} \sqrt{\frac{\omega_{\mathbf{k},\sigma}}{2M_{\beta}}} \left(a_{\mathbf{k},\sigma}\epsilon_{\mathbf{k},\sigma}^\beta e^{i\mathbf{k} \cdot \mathbf{R}_n} + h.c.\right).
\end{align}
Accordingly, we can rewrite the product of displacement and velocity operators as 
\begin{align}
\frac{u_m^\alpha \dot{u}_n^\beta }{N}
&= \frac{-i}{N} \sum_{\mathbf q,\mathbf q',\sigma,\sigma'} 
   \frac{1}{\sqrt{4 M_{\alpha} M_{\beta}\omega_{\mathbf q,\sigma}}}
   \sqrt{\omega_{\mathbf q',\sigma'}} \\
&\quad \times \Big[
   -a_{\mathbf q,\sigma} a_{\mathbf q',\sigma'}
      \epsilon^{\alpha}_{\mathbf q,\sigma}\epsilon^{\beta}_{\mathbf q',\sigma'}
      e^{i(\mathbf q\cdot \mathbf{R_m} + \mathbf q'\cdot \mathbf{R_n})} \notag \\
&\qquad -a^\dagger_{\mathbf q,\sigma} a_{\mathbf q',\sigma'}
      \epsilon^{\alpha *}_{\mathbf q,\sigma}\epsilon^{\beta}_{\mathbf q',\sigma'}
      e^{-i\mathbf q\cdot \mathbf R_m + i\mathbf q'\cdot \mathbf R_n} \notag \\
&\qquad +a_{\mathbf q,\sigma} a^\dagger_{\mathbf q',\sigma'}
      \epsilon^{\alpha}_{\mathbf q,\sigma}\epsilon^{\beta *}_{\mathbf q',\sigma'}
      e^{i\mathbf q\cdot \mathbf R_m - i\mathbf q'\cdot \mathbf R_n} \notag \\
&\qquad +a^\dagger_{\mathbf q,\sigma} a^\dagger_{\mathbf q',\sigma'}
      \epsilon^{\alpha *}_{\mathbf q,\sigma}\epsilon^{\beta *}_{\mathbf q',\sigma'}
      e^{-i(\mathbf q\cdot \mathbf R_m + \mathbf q'\cdot \mathbf R_n)}
   \Big]. \notag
\end{align}
Because the current operator in Eq.~\eqref{eq:current} contains an explicit dependence on $R_m-R_n$, the real–space sums leave only those terms whose phase factors depend on this difference. In practice, this requirement translates into the following selection rules for the four operator terms:
\begin{align*}
aa: && \bm{q}' = -\bm{q} = \mathbf{k}, 
&\qquad e^{i(\mathbf q\cdot\mathbf R_m+\mathbf q'\cdot\mathbf R_n)}
   \longrightarrow e^{-i\mathbf k\cdot(\mathbf R_m-\mathbf R_n)}, \\[0.5em]
a^\dagger a: && \bm{q}' = \bm{q} = \mathbf{k}, 
&\qquad e^{-i\mathbf q\cdot\mathbf R_m+i\mathbf q'\cdot\mathbf R_n}
   \longrightarrow e^{-i\mathbf k\cdot(\mathbf R_m-\mathbf R_n)}, \\[0.5em]
a a^\dagger: && \bm{q}' = \bm{q} = -\mathbf{k}, 
&\qquad e^{i\mathbf q\cdot\mathbf R_m-i\mathbf q'\cdot\mathbf R_n}
   \longrightarrow e^{-i\mathbf k\cdot(\mathbf R_m-\mathbf R_n)}, \\[0.5em]
a^\dagger a^\dagger: && -\bm{q}' = \bm{q} = \mathbf{k}, 
&\qquad e^{-i(\mathbf q\cdot\mathbf R_m+\mathbf q'\cdot\mathbf R_n)}
   \longrightarrow e^{-i\mathbf k\cdot(\mathbf R_m-\mathbf R_n)}.
\end{align*}
Next we consider the dynamical matrix
\begin{align}
D_{\alpha\beta}(\mathbf k)=\frac{1}{N\sqrt{M_\alpha M_\beta}}
\sum_{m,n}\Phi^{\alpha\beta}_{mn}e^{-i\mathbf k\cdot(\mathbf R_m-\mathbf R_n)},
\end{align}
which implies the following identity
\begin{align}
i\partial_{k_j}D_{\alpha\beta}(\mathbf k)
=\frac{1}{N\sqrt{M_\alpha M_\beta}}
\sum_{m,n}(\mathbf{R}_m-\mathbf{R}_n)_j\Phi^{\alpha\beta}_{mn}e^{-i\mathbf k\cdot(\mathbf R_m-\mathbf R_n)} .
\end{align}
Thus every term containing $(\mathbf{R}_m-\mathbf{R}_n)_je^{-i\mathbf k\cdot(\mathbf R_m-\mathbf R_n)}$ can be rewritten as $i\sqrt{M_\alpha M_\beta}\partial_{k_j}D_{\alpha\beta}(\mathbf k)$. Inserting these identities back into Eq.~\eqref{eq:current} and restricting the sums to wavevectors with $k_y>0$ (so that only forward–propagating phonon–polaritons contribute as in the setup), the current operator can be written entirely in terms of the dynamical matrix and its derivatives,
\begin{align}
J_i
=\frac{\hbar}{4V_c}\sum_{\mathbf k=k_y>0}\sum_{\sigma,\sigma'}
\frac{1}{\sqrt{\omega_{\mathbf k,\sigma}\omega_{\mathbf k,\sigma'}}}
\boldsymbol\epsilon_{\mathbf k,\sigma}^\dagger\left(\partial_{k_i}D(\mathbf k)\right)\boldsymbol\epsilon_{\mathbf k,\sigma'}
a^\dagger_{\mathbf k,\sigma}a_{\mathbf k,\sigma'}+\text{h.c.}
\label{eq:currentdynamical}
\end{align}
Substituting the identity
\begin{equation}
\boldsymbol\epsilon_{\mathbf k,\sigma}^\dagger
(\partial_{k_i}D)
\boldsymbol\epsilon_{\mathbf k,\sigma'}
= (\omega_{\mathbf k,\sigma}^2-\omega_{\mathbf k,\sigma'}^2)
\boldsymbol\epsilon_{\mathbf k,\sigma}^\dagger(\partial_{k_i}\boldsymbol\epsilon_{\mathbf k,\sigma'})
+ \delta_{\sigma\sigma'}\partial_{k_i}\left(\omega_{\mathbf k,\sigma}^2\right)
\label{eq:D-deriv-identity-all}
\end{equation}
back into Eq.~\eqref{eq:currentdynamical} yields
\begin{align}
J_i = \frac{\hbar}{2V_c}\sum_{\mathbf{k}>0}\Bigg[ &
\sum_{\sigma} \omega_{\mathbf{k},\sigma}\partial_{k_i}\omega_{\mathbf{k},\sigma}
a^\dagger_{\mathbf{k},\sigma} a_{\mathbf{k},\sigma} \quad +\label{eq:JgeneralSUPP}\\
&
\frac{1}{2}\sum_{\sigma\neq\sigma'} \sqrt{\frac{\omega_{\mathbf{k},\sigma'}}{\omega_{\mathbf{k},\sigma}}}
\big(\omega_{\mathbf{k},\sigma}^2-\omega_{\mathbf{k},\sigma'}^2\big)
\big(\epsilon^{\ast}_{\mathbf{k},\sigma}\cdot\partial_{k_i}\epsilon_{\mathbf{k},\sigma'}\big)
a^\dagger_{\mathbf{k},\sigma} a_{\mathbf{k},\sigma'} + h.c.
\Bigg]. \nonumber
\end{align}

\section*{First-order correction in the eigenvectors}

For completeness, we now also derive the first-order correction in the eigenvectors. Let $\{\epsilon_{\mathbf k,\sigma}\}$ be the orthonormal eigenvectors of the dynamical matrix fulfilling
\begin{equation}
    D(\mathbf k)\epsilon_{\mathbf k,\sigma}
    = \omega_{\mathbf k,\sigma}^{2}\epsilon_{\mathbf k,\sigma}, 
    \qquad
    \epsilon_{\mathbf k,\sigma}^\dagger \epsilon_{\mathbf k,\sigma'}=\delta_{\sigma\sigma'}.
    \label{eq:eigprob}
\end{equation}
Considering only forward-propagating wavevectors, $\{\mathbf k \mid k_y>0\}$, the quadratic Hamiltonian reads
\begin{equation}
    H = \sum_{\mathbf k, k_y>0}\sum_{\sigma,\sigma'}
    \epsilon_{\mathbf k,\sigma}^\dagger D(\mathbf k)\epsilon_{\mathbf k,\sigma'}
    a^{\dagger}_{\mathbf k,\sigma} a_{\mathbf k,\sigma'} ,
    \label{eq:H-epsDeps-forward}
\end{equation}
Since we study the response of phonons in a static magnetic field $\mathbf{B}$, we separate the quadratic Hamiltonian into an unperturbed harmonic term and a Zeeman-coupling term,
\begin{equation}
    H = H_0 + H_Z .
    \label{eq:H-split-forward}
\end{equation}
Let $\{\omega^{(0)}_{\mathbf k,\sigma},\epsilon^{(0)}_{\mathbf k,\sigma}\}$ be the eigenvalues and the eigenvectors of the unperturbed dynamical matrix $D^{(0)}(\mathbf k)$. Restricting to forward-propagating wavevectors $\{\mathbf k \mid k_y>0\}$, the harmonic Hamiltonian reads
\begin{equation}
    H_0
    = \sum_{\mathbf k,k_y>0}\sum_{\sigma}
      \hbar\omega^{(0)}_{\mathbf k,\sigma}
      \Big(a^{(0)\dagger}_{\mathbf k,\sigma} a^{(0)}_{\mathbf k,\sigma} + \tfrac{1}{2}\Big).
    \label{eq:H0-forward}
\end{equation}
The magnetic field now couples circular phonon-polariton modes via the phonon magnetic moment,
$\mu_{ph}$, as follows,
\begin{equation}
    H_Z
    = -\mu_{\rm ph}\sum_{\mathbf k, k_y>0}\sum_{\sigma,\sigma'}
    \sqrt{\frac{\omega^{(0)}_{\mathbf k,\sigma'}}{\omega^{(0)}_{\mathbf k,\sigma}}}
    \mathbf B\cdot\Big[i\epsilon^{(0)*}_{\mathbf k,\sigma}\times\epsilon^{(0)}_{\mathbf k,\sigma'}\Big]
    a^{(0)\dagger}_{\mathbf k,\sigma} a^{(0)}_{\mathbf k,\sigma'} .
    \label{eq:HZ-forward}
\end{equation}
Treating $H_Z$ as a weak perturbation, we expand the eigenvectors as
\begin{equation}
    \epsilon_{\mathbf k,\sigma}
    = \epsilon^{(0)}_{\mathbf k,\sigma}
         + \epsilon^{(1)}_{\mathbf k,\sigma}
         + \mathcal O(B^2),
\end{equation}
Since we work in the circular eigenvector basis (left- and right-circular modes),
the Zeeman term does not couple the two eigenvectors directly. This allows us to
apply nondegenerate perturbation theory, where the first-order correction is given by
\begin{equation}
    \epsilon^{(1)}_{\mathbf k,\sigma'}
    = \sum_{\sigma''\neq\sigma'}
      \frac{\Delta^{(Z)}_{\sigma''\sigma'}(\mathbf k)}
           {\omega^{(0)2}_{\mathbf k,\sigma'}-\omega^{(0)2}_{\mathbf k,\sigma''}}
      \epsilon^{(0)}_{\mathbf k,\sigma''},
    \qquad
    \Delta^{(Z)}_{\sigma\sigma'}(\mathbf k)
    := \langle\epsilon^{(0)}_{\mathbf k,\sigma} | H_Z | \epsilon^{(0)}_{\mathbf k,\sigma'}\rangle .
    \label{eq:eps1}
\end{equation}

\end{document}